\documentclass[11pt,english]{article}

\usepackage[koi8-r]{inputenc}
\usepackage[russian]{babel}
\usepackage{epsfig}
\usepackage{amssymb}

\begin{document}
\selectlanguage{english}

\title{\bf Study of the process $e^+e^-\to\eta\gamma\to 7\gamma$
in the energy range $\sqrt{s}$ = 1.07 -- 2 GeV }
\author{M.N.Achasov,
A.Yu.Barnyakov,
K.I.Beloborodov,
A.V.Berdyugin\footnote{email:berdugin@inp.nsk.su}, \\
A.G.Bogdanchikov,
A.A.Botov,
V.S.Denisov,
T.V.Dimova, \\
V.P.Druzhinin,
L.B.Fomin,
A.G.Kharlamov, 
L.V.Kardapoltsev, \\
A.N.Kyrpotin,
I.A.Koop,    
A.A.Korol,   
D.P.Kovrizhin,
A.P.Kryukov, \\
A.S.Kupich,   
N.A.Melnikova,        
N.Yu.Muchnoi,
A.E.Obrazovsky, \\
E.V.Pakhtusova,
E.A.Perevedentsev,
K.V.Pugachev,  
Yu.A.Rogovsky, \\
S.I.Serednyakov,
Z.K.Silagadze,  
I.K.Surin,
M.V.Timoshenko, \\
Yu.V.Usov,
V.N.Zhabin,     
V.V.Zhulanov,
I.M.Zemlyansky, \\
Yu.M.Shatunov,
D.A.Shtol,
A.N.Shukaev,
E.A.Eminov \\
\it Budker Institute of Nuclear Physics, SB RAS, Novosibirsk, 630090, Russia\\
\it Novosibirsk State University, Novosibirsk, 630090, Russia\\
}

\date{}
\maketitle

\begin{abstract}
The $e^+e^-\to\eta\gamma$ cross section is measured in the
center-of-mass energy range from 1.07 to 2.00 GeV in the decay
channel $\eta\to 3\pi^0$, $\pi^0\to\gamma\gamma$. The data set 
with an integrated luminosity of 242 pb$^{-1}$ accumulated in the
experiment with the SND detector at the VEPP-2000 $e^+e^-$ collider
is analyzed.
\end{abstract}

\section{Introduction}
Radiative decays are one of the best tools for study of the internal
structure of hadrons. For light vector mesons, these decays have been
studied for more than 50 years. The decay probabilities of the $\rho$,
$\omega$, and $\phi$ resonances into the final state $\eta\gamma$ are
currently measured with an accuracy of 7\%, 9\%, and 2\%, respectively.
Moreover, for $\rho$ and $\omega$ mesons, the uncertainty
is still determined by statistics. The most accurate measurements of
$\rho$, $\omega$, and $\phi\to\eta\gamma$ decays were made in the
SND~\cite{SNDetagam2007} and CMD-2~\cite{CMDetagam2001} experiments at 
the VEPP-2M $e^+e^-$ collider.

In $e^+e^-$ experiments, the directly measured quantity is the cross section
for the process $e^+e^-\to\eta\gamma$. The decay probabilities can be
determined from the fit to the cross section with the sum of the
contributions of vector resonances. From the analysis of the VEPP-2M
data~\cite{SNDetagam2007}, it was found that the model uncertainty 
of the $\rho$, $\omega$, and $\phi\to \eta\gamma$ branching fractions
associated with the uncertainty of the contributions of excited vector states
reach several percent. To eliminate this uncertainty, it is required, in 
particular, to measure the $e^+e^-\to\eta\gamma$ cross section
at center-of-mass energies $\sqrt{s}$ at least up to 2 GeV.

The measurement at $\sqrt{s} = 1.05$--2 GeV is important in itself.
From it one can extract the probabilities of radiative decays of excited
vector mesons $\rho(1450)$, $\rho(1700)$ and $\phi(1680)$. In this energy
region, in addition to the conventional vector $q\bar{q}$ states, the 
production of exotic hybrid (quark-antiquark-gluon) mesons is possible. 
Since hybrid states can mix with two-quark states, their identification is
a complex experimental task requiring a detailed analysis of all available
decay modes. Radiative decays, whose probabilities are relatively well 
predicted within the framework of the quark model, may turn out to be the key
to identifying vector hybrid states~\cite{SNDphi2etapg2003}.

In this paper, we present the measurement of the $e^+e^-\to\eta\gamma$ 
cross section in the energy range $\sqrt{s}=1.07$--2.00 GeV in the 
experiment with the SND detector at the VEPP-2000 $e^+e^-$ 
collider~\cite{VEPP2000}. We use statistics with an integrated luminosity
of about 242 pb$^{-1}$ accumulated from 2010 to 2021. The results of 
the measurement of the $e^+e^-\to\eta\gamma$ process in this energy range,
obtained by SND based on 2010--2012 data with an integrated luminosity 
of about 36 pb$^{-1}$, were published in Ref.~\cite{SNDetagam2014}. 
Since this publication, in the SND, CMD-3, and BABAR experiments,
the cross sections for the background processes 
$e^+e^-\to K_SK_L\pi^0$~\cite{BABARkkpi2017,SNDkkpi2018} and 
$e^+e^ -\to K_SK_L\pi^0\pi^0$~\cite{BABARkkpi2017}
were refined, while the cross sections for the processes
$e^+e^-\to\eta\eta\gamma$~\cite{SNDomegaeta2019,CMDkketa2019},
$e^+e^-\to\omega\eta\pi^0$~\cite{SNDomegaetapi2016,BABARomegaetapi2018,BABARomegaetapi2021},
and $e^+e^-\to K_SK_L\eta$~\cite{BABARkkpi2017} 
were measured for the first time. These data are used in the new analysis.

\section{Detector and experiment}
During the experiments, the energy interval 1.05--2.00 GeV was scanned
several times with a step of 20-25 MeV. In this analysis, due to limited
statistics, we present as a result the cross section values
averaged over 14 energy intervals listed in Table~\ref{tabl}.

A detailed description of the SND detector is given in Refs.~\cite{SND}.
It is a non-magnetic detector, the main part of which
is a three-layer spherical electromagnetic calorimeter based on NaI(Tl)
crystals. The solid angle of the calorimeter is 95\% of 4$\pi$. Its energy
resolution for photons is $\sigma_E/E=4.2\%/\sqrt[4]{E({\rm \mbox{GeV}})}$,
and its angular resolution is about $1.5^\circ$. The angles and
production vertex of charged particles are measured in a tracking system 
consisting of a nine-layer drift chamber and a proportional chamber with 
signal readout with cathode strips. The solid angle of the tracking system
is 94\% of 4$\pi$.

The main $\eta$ meson decay modes are $2\gamma$ (39\%), $3\pi^0$ (33\%) and
$\pi^+\pi^-\pi^0$ (23\%). Background from the processes $e^+e^-\to 3\gamma$
and $e^+e^-\to \pi^+\pi^-2\pi^0$, which significantly exceeds the effect in
the energy range 1.07--2.00 GeV makes it difficult to use the
$\eta\to 2\gamma$ and $\eta\to \pi^+\pi^-\pi^0$ decay modes. In this paper,
the $e^+e^-\to\eta\gamma$ process is studied in the 
$\eta\to 3\pi^0~,~\pi^0\to 2\gamma$ decay channel, which has seven photons in
the final state. Since the final state for the process under study does not 
contain charged particles, the process without charged particles 
$e^+e^-\to \gamma\gamma$ is also chosen for normalization. As a result of 
the normalization, the systematic uncertainties associated with the 
event selection in the first level trigger, as well as the uncertainties 
arising due to superimposing of beam-generated background charged tracks on  
events under study, are canceled. The accuracy of the luminosity measurement
using the $e^+e^-\to \gamma\gamma$ process is 2.2\%~\cite{SNDomegapi2013}.

\section{Selection conditions}
The background processes are 
$e^+e^-\to\pi^0\pi^0\gamma$,
$e^+e^-\to\eta\pi^0\gamma$, 
$e^+e^- \to\eta\eta\gamma$,
$e^+e^-\to\omega\pi^0\pi^0$, and 
$e^+e^-\to\omega\eta\pi^0$ with decays
$\omega\to\pi^0\gamma$, $\eta\to 3\pi^0$ and $\eta\to\gamma\gamma$. 
The processes with neutral kaons 
$e^+e^-\to K_SK_L(\gamma)$,
$e^+e^-\to K_SK_L\pi^0$, 
$e^+e^- \to K_SK_L\pi^0\pi^0$, and 
$e^+e^-\to K_SK_L\eta$ with $K_S\to 2\pi^0$ decay 
also contribute to background.

Of the above processes, only $e^+e^-\to\omega\pi^0\pi^0$ and
$\omega\eta\pi^0$ have seven photons in the final state. In processes with
the $K_L$ meson, additional photons can be reconstructed due to the $K_L$ 
nuclear interaction in the calorimeter or its decay. 
Also, additional photons are arised from splitting of electromagnetic showers
in the calorimeter, emission of photons by the initial particles at a large 
angle, and superimposing of beam-generated background.

The selection of events is carried out in two stages. First, events are
selected, in which seven or more photons are detected and there are no
charged particles, with the following conditions on the total energy
deposition in the calorimeter $E_{\rm tot}$ and the total event momentum 
$P_{\rm tot}$ calculated using the energy depositions in the calorimeter 
crystals
\begin{equation}
0.7 < E_{\rm tot}/\sqrt{s} < 1.2,~P_{\rm tot}/\sqrt{s} < 0.3,~
E_{\rm tot}/\sqrt{s} - P_{\rm tot}/\sqrt{s} > 0.7.
\end{equation}

For selected events, a kinematic fit is performed using the
measured photon angles and energies, energy-momentum conservation laws, and
assumptions about the presence of intermediate $\pi^0$ mesons. As a result
of the fit, the photon energies are refined and $\chi^2$ is
calculated for the used kinematic hypothesis. The two hypotheses are tested:
\begin{itemize}
\item[ ] $e^+e^-\to 3\pi^0\gamma$ ($\chi^2_{3\pi^0\gamma}$),
\item[ ] $e^+e^-\to\pi^0\pi^0\gamma$ ($\chi^2_{\pi^0\pi^0\gamma}$).
\end{itemize}
In the $e^+e^-\to 3\pi^0\gamma$ hypothesis, the photon with the maximum
energy is chosen as the recoil photon. The $\pi^0$ candidates are formed from
the remaining six photons. If there are more photons in an event compared to
the requirement of the hypothesis, all possible five (seven)-photon
combinations are tested and the combination with the minimum value of
$\chi^2_{\pi^0\pi^0\gamma}$ ($\chi ^2_{3 \pi^0\gamma}$) is selected.

Further selection of events is carried out according to the following
conditions:
\begin{equation}
\chi^2_{3\pi^0\gamma}<50,~\chi^2_{\pi^0\pi^0\gamma}>20.
\end{equation}

\begin{figure}[p]
\includegraphics[width=0.85\textwidth]{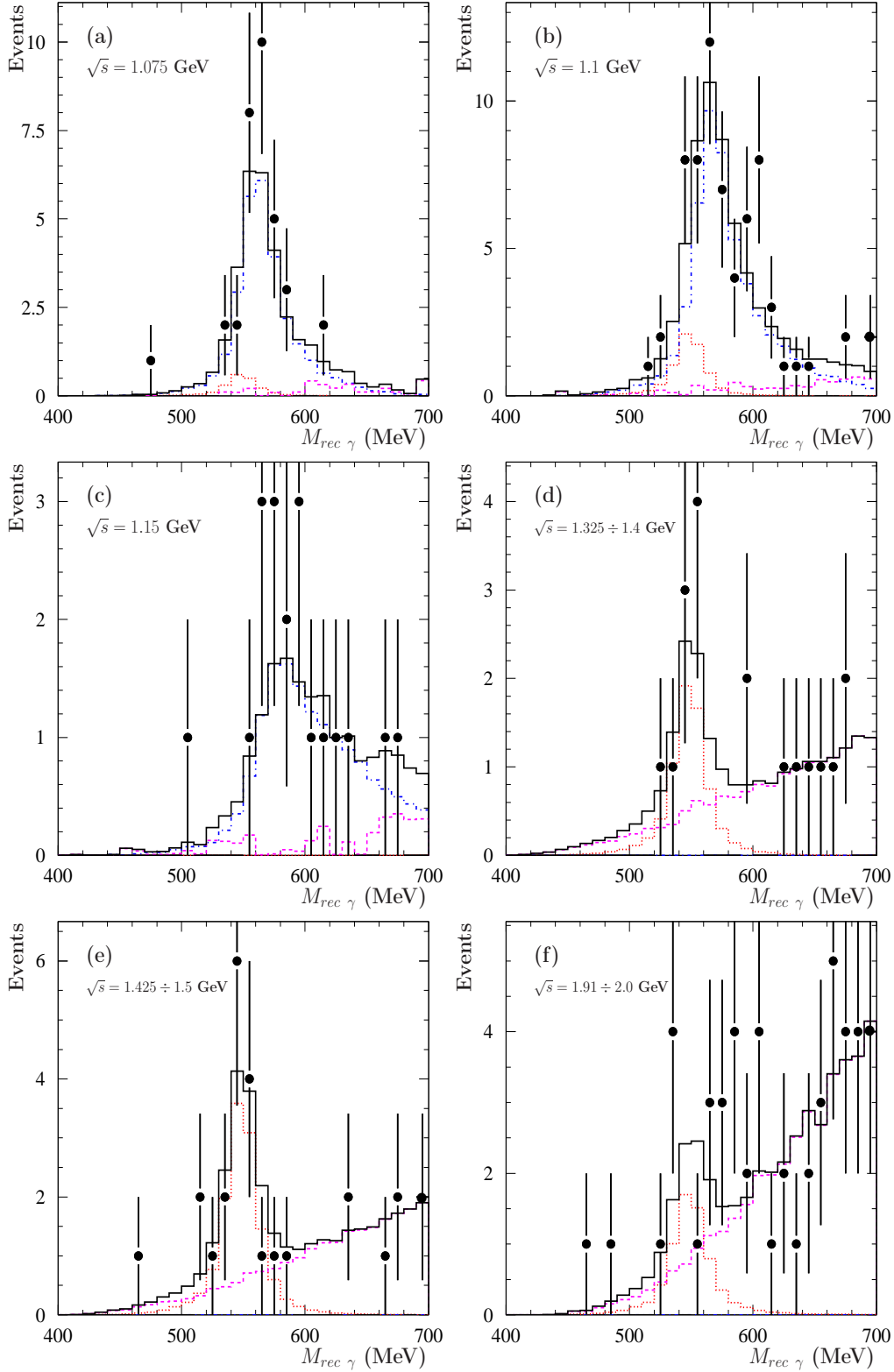}
\caption{ \label{rmg}
The $M_{\rm rec}$ distributions for six energy intervals. The points with 
errors bar represent data, the solid histogram is the result of the fit 
described in the text. The dotted histogram is the fitted contribution of the
$e^+e^-\to\eta\gamma$ process, the dash-dotted histogram is the calculated 
$e^+e^-\to\phi \gamma\to\eta\gamma\gamma$ contribution, 
dashed histogram is the sum of all other background processes.
}
\end{figure}

For selected events, the distribution of the invariant mass $M_{\rm rec}$ 
recoiling against the photon in the $e^+e^-\to 3\pi^0\gamma$ hypothesis is 
analyzed. These distributions in the range $400 < M_{\rm rec} < 700$ MeV for
six energy intervals are shown in Fig.~\ref{rmg}.

At energies below 1.3 GeV, a significant contribution to the $M_{\rm rec}$
distribution comes from the process of ``radiative return'' to the
$\phi$-meson resonance 
$e^+e^-\to\phi\gamma_{\rm ISR}\to\eta\gamma\gamma_{\rm ISR}$, 
in which the additional photon $\gamma_{\rm ISR}$ is emitted from the initial
state predominantly at a small angle to the beam axis. We consider this 
process as background. Its contribution with the condition on the mass 
recoiling against $\gamma_{\rm ISR}$: $\sqrt{s^\prime} < 1.03$ GeV, 
is calculated by Monte-Carlo (MC) simulation using data
on the $e^+e^-\to\eta\gamma$ cross section at energies below 1.03
GeV~\cite{SNDetagam2007}. The calculated $M_{\rm rec}$ spectrum for the
$e^+e^-\to\phi\gamma_{\rm ISR}$ process is shown in Fig.~\ref{rmg}, and the
expected number of events with $400 < M_{\rm rec} < 700$ MeV
is listed in Table~\ref{tabl}.

The contribution of other background processes is calculated using
data on the measured cross sections for
$e^+e^-\to\pi^0\pi^0\gamma$~\cite{SNDpi0pi0g2016}, 
$e^+e^-\to\eta\pi^0\gamma$~\cite{SNDomegaeta2019,SNDetapi0g2020},
$e^+e^-\to\eta\eta\gamma$~\cite{SNDomegaeta2019,CMDkketa2019},
$e^+e^-\to \omega\pi^+\pi^-$~\cite{CMDomegapipi2000,BABARomegapipi2007}, 
$e^+e^-\to\omega\eta\pi^0$~\cite{SNDomegaetapi2016,BABARomegaetapi2018,BABARomegaetapi2021},
$e^+e^-\to K_SK_L(\gamma)$~\cite{BABARkskl2014},
$e^+e^-\to K_SK_L\pi^0$~\cite{BABARkkpi2017,SNDkkpi2018},
$e^+e^-\to K_SK_L\pi^0\pi^0$~\cite{BABARkkpi2017}
and $e^+e^-\to K_SK_L\eta$~\cite{BABARkkpi2017}.
For the process $e^+e^-\to \omega\pi^0\pi^0$, the isotopic relation
$\sigma(\omega\pi^+\pi^-)=2\sigma(\omega\pi^ 0\pi^0)$ is used. Radiative
corrections~\cite{radcor} are taken into account when calculating the
background. This is especially important for the $e^+e^-\to K_S K_L(\gamma)$
process, which is dominated by radiative return to the $\phi$ meson:
$e^+e^-\to\phi\gamma\to K_SK_L\gamma$.

For the energy range above 1.6 GeV, the cross sections for many background
processes are known with an accuracy of about 25\%. The cross section of the
$e^+e^-\to\omega\eta\pi^0$ process measured in the SND and BABAR
experiments differs by a factor of 2. Below 1.2 GeV, the dominant
background source is the $e^+e^-\to K_SK_L(\gamma)$ process. The accuracy
of its estimation is determined by the quality of MC simulation of
$K_L$ nuclear interaction in the calorimeter. Therefore,
the mass interval $700 < M_{\rm rec} < 1100$ MeV is also analyzed, where
only background processes are expected to contribute.

The $M_{\rm rec}$ distributions in the range $400 < M_{\rm rec} < 1100$ MeV
are fitted by a sum of the contributions of the process under study 
$e^+e^-\to\eta\gamma$ and background processes:
\begin{eqnarray}
P(M_{\rm rec}) = N_{\eta\gamma}P_{\eta\gamma}(M_{\rm rec}) +
\alpha_{\rm bkg}P_{\rm bkg}(M_{\rm rec}) + P_{\phi\gamma}(M_{\rm rec}). \label{Ffit}
\end{eqnarray}
Here $P_{\eta\gamma}$ is the signal distribution normalized to unity,
$P_{\phi\gamma}$ is the calculated spectrum for the process
$e^+e^-\to\phi\gamma_{\ rm ISR}\to\eta\gamma\gamma_{\rm ISR}$, and 
$P_{\rm bkg}$ is the calculated spectrum for other background processes. 
The free fit parameters are the number of signal events 
$N_{\eta\gamma}$ and the scale factor for the background $\alpha_{\rm bkg}$.
Below 1.4 GeV, statistics do not allow to determine $\alpha_{\rm bkg}$ with
the required accuracy for each interval. Therefore,
to determine the background, the combined $M_{\rm rec}$ distributions are
fitted in the ranges $\sqrt{s}<1.225$ GeV and $1.225<\sqrt{s}<1.4$ GeV. 
The resulting $\alpha_{\rm bkg}$ values with their error are then
used in the fit for individual intervals.

The shape of the distribution for $M_{\rm rec}$ was checked according to the data
collected near the $\phi-$resonance. The simulation agrees with experiment. For the
purposes of these statistics the shape of the distribution for $M_{\rm rec}$
does not need to be amended. 

The obtained numbers of events of the signal and background processes, as
well as the values of the coefficient $\alpha_{\rm bkg}$ for different
energy intervals, are listed in Table~\ref{tabl}.

\section{Detection efficiency}
\begin{figure}
\includegraphics[width=0.95\textwidth]{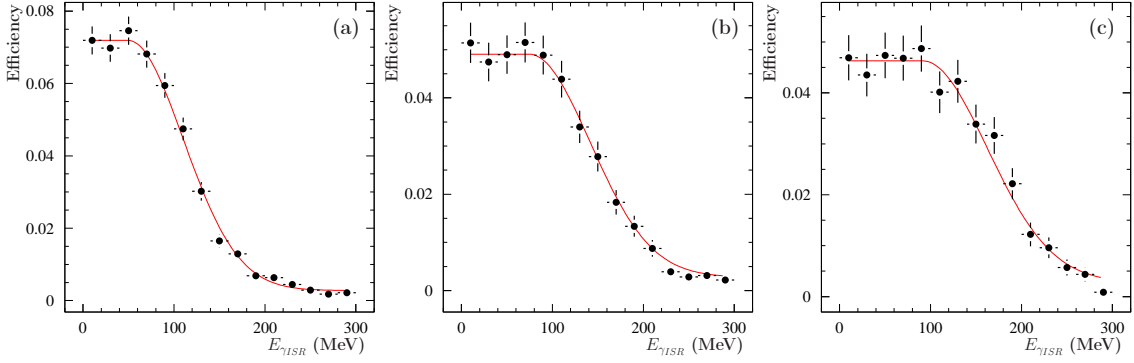}
\caption{ \label{eff}
The dependence of the detection efficiency for 
$e^+e^-\to\eta\gamma(\gamma)$ events on the energy of the additional photon
emitted by the initial particles at $\sqrt{s} = 1.15$ GeV (a), 1.6 GeV
(b), and 1.9 GeV (c). The points with error bars are obtained using 
MC simulation, the curve shows the result fit to the 
$\varepsilon(E_{\gamma_{\rm ISR}})$ dependence with a smooth function.
}
\end{figure}

The signal detection efficiency is
determined by MC simulation, which take into account radiative
corrections to the initial state~\cite{radcor}, in particular, the emission
of additional photons. The angular distribution of these photons is
generated according to Ref.~\cite{BM}. Figure~\ref{eff} shows the dependence
of the detection efficiency $\varepsilon(\sqrt{s},E_{\gamma_{\rm ISR}})$ on 
the energy $E_{\gamma_{\rm ISR}}$ of the photon radiated from the initial
state for three values of the center-of-mass energy.

The values of the detection efficiency at $E_{\gamma_{\rm ISR}}=0$, averaged
over the energy intervals, are listed in Table~\ref{tabl}.

\section{Cross section parametrization}
In the framework of the vector meson dominance model, the cross section of
the $e^+e^-\to\eta\gamma$ process can be written as:
\begin{equation}
\label{parcrs1}
\sigma_{\eta\gamma}(\sqrt{s}) = \left( \frac{k_\gamma(\sqrt{s})}{\sqrt{s}} \right)^3
\left| \sum\limits_{V=\rho,~\omega,~\phi,{\ldots} }
A_{V}(\sqrt{s})\right|^2,
\end{equation}
\begin{equation}
\label{parcrs2}
A_V(\sqrt{s}) = \frac{m_V \Gamma_V(m_V) e^{i\varphi_V}}{D_V(\sqrt{s})} \sqrt{ \frac{m^3_V}
{k_\gamma(m_V)^3} \sigma_{V\eta\gamma}},
\end{equation}
\begin{equation}
\label{parcrs3}
D_V(\sqrt{s}) = m^2_V - s - i \sqrt{s}\Gamma_V(\sqrt{s}),\;\;\;
k_\gamma(\sqrt{s}) = \frac{\sqrt{s}}{2} \left( 1 - \frac{m^2_{\eta}}{s} \right),
\end{equation}
where the summation is over all vector resonances $V$ that contribute to the
cross section, $m_V$ and $\Gamma_V(\sqrt{s})$ are the mass of the resonance
and its total width, 
$\sigma_{V\eta\gamma } =(12\pi/m_V^2)B(V\to e^+e^-)B(V\to \eta\gamma)$
is the cross section of the process $e^+e^-\to V \to\eta\gamma$ for 
$\sqrt{s}=m_V$, $B(V\to e^+e^-)$ and $B(V\to \eta\gamma)$
are the branching fractions of the corresponding decays , $\varphi_V$ 
are the phases of the vector resonance amplitudes ($\varphi_{\rho} \equiv 0$).
In addition to the resonances $\rho$, $\omega$, and $\phi$, the sum in 
Eq.~(\ref{parcrs1}) includes all their excited states. For $\rho$,
$\omega$, and $\phi$, when calculating the energy dependence of the widths,
the main decay modes are taken into account. For excited
resonances, the widths were assumed to be independent of the energy.

\section{Fit to data and obtaining the Born cross section\label{fit}}
The visible cross section of the $e^+e^-\to \eta\gamma$ process is related
to the Born cross section ($\sigma(\sqrt{s})$), which must be determined
from experiment, by the following formula:
\begin{equation}
\label{viscrs}
\sigma_{vis}(\sqrt{s}) = \int\limits_{0}^{x_{max}}
\varepsilon \left( \sqrt{s},\frac{x\sqrt{s}}{2} \right) 
F\left( x,\sqrt{s} \right) 
\sigma \left( \sqrt{s(1-x)} \right) dx~,
\end{equation}
where $F(x,\sqrt{s})$ is a function describing the distribution of the
energy fraction $x=2E_{\gamma_{\rm ISR}}/\sqrt{s}$ taken away by photons
emitted from the initial state~\cite{radcor}. The value of $x_{max}$ is 
determined by the condition $\sqrt{s^\prime}=\sqrt{s(1-x_{max})} < 1.03$ GeV,
which is used to separate the processes $e^+e^-\to\eta\gamma(\gamma)$ and 
$e^+e^-\to \phi\gamma$.
The expression (\ref{viscrs}) can be rewritten as:
\begin{equation}
\label{viscrs1}
\sigma_{vis}(\sqrt{s}) = \varepsilon_0(\sqrt{s})\,\sigma(\sqrt{s})\,(1+\delta(\sqrt{s}))~,
\end{equation}
where the detection efficiency $\varepsilon_0(\sqrt{s})$ and the radiative
correction $\delta(\sqrt{s})$ are defined as follows:
\begin{equation}
\varepsilon_0(\sqrt{s}) \equiv \varepsilon(\sqrt{s},0),
\end{equation}
\begin{equation}
\delta(\sqrt{s}) = \frac{\int\limits_{0}^{x_{max}}
\varepsilon \left( \sqrt{s},\frac{x\sqrt{s}}{2} \right) ~F(x,\sqrt{s})~
\sigma \left( \sqrt{(1-x)s} \right) dx }{\varepsilon_r(\sqrt{s},0)\sigma(\sqrt{s})}-1.
\end{equation}
Technically, the Born cross section is found as follows. The energy dependence
of the measured visible cross section
$\sigma_{vis}(\sqrt{s_i}) = N_{\eta\gamma,i}/IL_i$, where $i$ is the energy
interval number, is fitted by Eq.~(\ref{viscrs}). For the parametrization of
the Born cross section, some theoretical model is used that describes the
experimental data well. Using the obtained parameters of the theoretical
model, the radiative correction $\delta(\sqrt{s_i})$ is determined, and then
the experimental Born cross section~$\sigma(\sqrt{s_i})$ is calculated using
the formula (\ref{viscrs1}).

In the fit to the cross section, the Particle Data Group (PDG) values of the 
$\rho$, $\omega$, and $\phi$ parameters~\cite{pdg} are used. The phases $\rho$,
$\omega$ and $\phi$ are chosen according to the prediction of the quark
model: $\varphi_{\omega}=\varphi_{\rho}$, 
$\varphi_{\phi}=\varphi_{\rho}+180^\circ$. As already mentioned, at energies
above 1 GeV, all five known excited vector resonances $\omega(1420)$,
$\rho(1450)$, $\omega(1650)$, $\phi(1680)$, and $\rho(1700)$ contribute to 
the $e^+e^-\to \eta\gamma$ cross section. 
It is impossible to separate the contributions of these resonances by
fitting the cross section $e^+e^-\to \eta\gamma$.
However, the problem can be significantly simplified using the fact that 
the resonances are divided into two groups with close masses 
($\omega(1420)$, $\rho(1450)$) and ($\omega(1650)$, $\phi(1680)$, and 
$\rho(1700)$). With the available limited statistics, we can use a model with
two effective resonances $\rho^{\prime}$ and $\phi^\prime$ with masses and 
widths equal to the PDG values for $\rho(1450)$ and $\phi( 1680)$. This choice
of resonances is consistent with the predictions of the quark 
model~\cite{thpred}, in which the decay widths $\rho(1450)\to\eta\gamma$ and
$\phi(1680)\to\eta\gamma$ are at least an order order of magnitude
larger than the widths for the other three excited states.

\begin{figure}
\begin{center}
\includegraphics[width=0.5\textwidth]{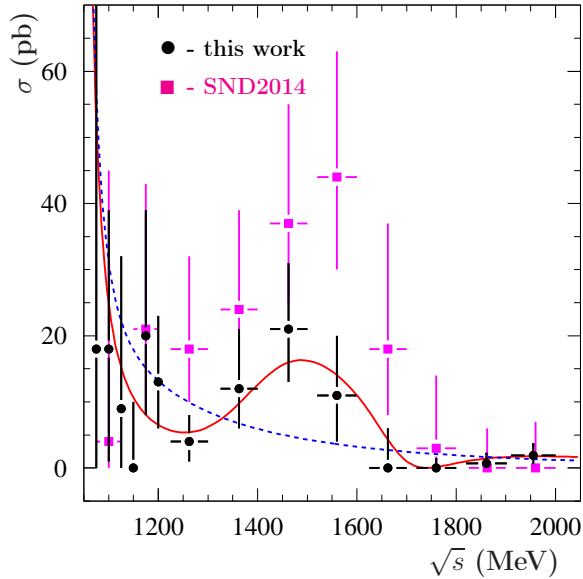}
\caption{ \label{fig:crs}
The cross section for the process $e^+e^-\to\eta\gamma$ measured in
this work in comparison with the cross section obtained earlier in 
Ref.~\cite{SNDetagam2014}. The dashed curve is the result of the fit
taking into account the contributions of only $\rho$, $\omega$, and $\phi$
mesons. The solid curve is the result of fit with the additional
contribution of two excited vector resonances.
}
\end{center}
\end{figure}

The free fit parameters are the cross sections
$\sigma_{\rho^{\prime}\eta\gamma}$ and $\sigma_{\phi^\prime\eta\gamma}$, and
the phases $\varphi_{\rho^\prime} $ and $\varphi_{\phi^{\prime}}$. The
resulting fitted curve is shown in Fig.~\ref{fig:crs} together with
the values of the Born cross section calculated using Eq.~(\ref{viscrs1})
The numerical values of the Born cross section and the radiative
correction are listed in Table~\ref{tabl}.

For the cross sections at the resonance maxima, the following values
are obtained:
\begin{eqnarray}
\sigma_{\rho^{\prime}\eta\gamma} = 16^{+15}_{-10} \pm 2\mbox{ pb},\nonumber \\
\sigma_{\phi^{\prime}\eta\gamma} = 14^{+14}_{-10} \pm 2\mbox{ pb}. \label{res1}
\end{eqnarray}
The first of the quoted  errors is statistical, the second is systematic.

It should be noted that small statistics do not allow us to discard
the model without excited resonances. The result of the fit in this hypothesis
is also shown in Fig.~\ref{fig:crs}. For it, $\chi^2/\nu =11.4/14$, where 
$\nu$ is the number of degrees of freedom, versus $\chi^2/\nu = 4.7/10$ in 
the model with two excited resonances. It should be noted that
the model with only one excited resonance cannot describe the cross
section dip at $\sqrt{s} = 1.75$ GeV.
\begin{table}[p]
\begin{center}
\caption{\label{tabl}	
The energy interval ($\sqrt{s}$), 
integrated luminosity ($IL$), 
number of events of the process $e^+e^-\to\phi\gamma\to\eta\gamma\gamma$
($N_{\phi\gamma}$) in the range $400 < M_{\rm rec} < 700$ MeV,
number of events of other background processes ($N_{\rm bkg}$) in 
the range $400 < M_{\rm rec} < 700$ MeV, 
background scale factor  ($\alpha_{\rm bkg}$), 
detection efficiency ($\varepsilon_0$), 
number of $e^+e^-\to\eta\gamma$ events ($N_{\eta\gamma }$), 
radiative correction factor ($1+\delta$), 
Born cross section for the process $e^+e^-\to\eta\gamma$ $\sigma$ ($\sigma$).
The first error in the cross section is statistical, the second is systematic.}
\begin{tabular}{cccccccc}
$\sqrt{s}$ (GeV) &
$L$ (pb$^{-1}$) &
$N_{\phi\gamma}$ &
$N_{\rm bkg} (\alpha_{\rm bkg})$ &
$\varepsilon_0$ (\%) &
$N_{\eta\gamma}$ &
$\delta+1$ &
$\sigma$ (pb)
\\ \hline \hline
 1.075        & 1.10 & 28 &   2.9 ($1.36\pm 0.27$) & 8.1 & $2.0^{+7.0}_{-2.0}$ &  $1.26 \pm 0.04$ & $  18^{+ 63}_{- 18} \pm 1$     \\
 1.100        & 3.38 & 51 &   6.4 ($1.36\pm 0.27$) & 8.0 & $7.0^{+8.1}_{-6.5}$ &  $1.43 \pm 0.12$ & $  18^{+ 21}_{- 17} \pm 1$     \\
 1.125        & 1.32 & 11 &   1.4 ($1.36\pm 0.27$) & 8.2 & $1.5^{+3.7}_{-1.5}$ &  $1.48 \pm 0.19$ & $   9^{+ 23}_{-  9} \pm 1$     \\
 1.150        & 3.21 & 15 &   2.8 ($1.36\pm 0.27$) & 8.1 & $0.0^{+3.8}$        &  $1.44 \pm 0.22$ & $   0^{+ 10}        \pm 0.1$   \\
 1.175        & 1.73 &  4 &   1.1 ($1.36\pm 0.27$) & 7.9 & $3.6^{+3.5}_{-2.2}$ &  $1.35 \pm 0.20$ & $  20^{+ 19}_{- 12} \pm 2$     \\
 1.200        & 4.30 &  4 &   2.2 ($1.36\pm 0.27$) & 7.7 & $5.3^{+4.2}_{-2.9}$ &  $1.25 \pm 0.16$ & $  13^{+ 10}_{-  7} \pm 1$     \\
 1.225--1.300 & 21.0 &  5 &  20   ($1.35\pm 0.15$) & 7.0 & $5.9^{+5.4}_{-3.9}$ &  $1.01 \pm 0.01$ & $   4^{+  4}_{-  3} \pm 0.2$   \\
 1.325--1.400 & 10.0 &  1 &  16   ($1.35\pm 0.15$) & 6.6 & $6.9^{+5.1}_{-3.7}$ &  $0.90 \pm 0.08$ & $  12^{+  9}_{-  6} \pm 1$     \\
 1.425--1.500 & 11.0 &  0 &  22   ($1.01\pm 0.11$) & 6.3 & $13.0^{+6.6}_{-5.2}$&  $0.91 \pm 0.07$ & $  21^{+ 10}_{-  8} \pm 2$     \\
 1.520--1.600 & 11.3 &  0 &  34   ($1.12\pm 0.08$) & 6.0 & $6.8^{+5.9}_{-4.4}$ &  $0.95 \pm 0.03$ & $  11^{+  9}_{-  7} \pm 0.4$   \\
 1.625--1.700 & 12.4 &  0 &  58   ($1.28\pm 0.07$) & 5.6 & $0.0^{+4.9}$        &  $1.18 \pm 0.20$ & $   0^{+  6}        \pm 0.3$   \\
 1.720--1.800 & 15.0 &  0 &  25   ($1.13\pm 0.08$) & 5.4 & $0.0^{+3.8}$        &  $2.94 \pm 1.94$ & $   0^{+1.5}        \pm 0.2$   \\
 1.820--1.902 & 63.5 &  0 &  43   ($1.01\pm 0.05$) & 4.9 & $1.9^{+4.6}_{-1.9}$ &  $0.92 \pm 0.06$ & $ 0.7^{+1.6}_{-0.7} \pm 0.1$   \\
 1.910--2.000 & 83.2 &  0 &  38   ($0.97\pm 0.05$) & 4.6 & $6.7^{+6.3}_{-4.7}$ &  $0.94 \pm 0.05$ & $ 1.9^{+1.8}_{-1.3} \pm 0.1$   \\
\end{tabular}
\end{center}
\end{table}

\section{Systematic measurement errors}
The systematic uncertainty on the measured cross section includes
uncertainties in determining the detection efficiency, measuring the 
integrated luminosity, as well as the model error in calculation the radiative
correction. To estimate the systematic error in the detection efficiency, we
study the stability of the result on the cross section under wide
variation of the selection criteria, in particular, the conditions on 
$\chi^2$ of the kinematic fits. An analysis is also carried out with the 
requirement that exactly seven photons be detected in an event, as in
Ref.~\cite{SNDetagam2014}. At the current level of statistical accuracy, no
change in the result on the cross section is found. In addition, for a
numerical estimate of the uncertainty in the detection efficiency, one can 
use the results of the study of the difference in the detector response 
between data and simulation for the five-photon events performed in
Ref.~\cite{SNDomegapi2013}. For the current analysis, we use the sum of the
correction from~\cite{SNDomegapi2013} and its error (3\%) as an estimate of
the systematic uncertainty associated with selection conditions. The systematic
uncertainty due to the difference between data and simulation in the
photon conversion probability before the track system is 1.3\%.

The systematic uncertainty associated with normalization to luminosity is
2.2\%. The model error in the calculation of the radiative correction is
determined from the difference between the values obtained for the
models with and without the use of $\rho^\prime$ and $\phi^{\prime}$
excited states. The total systematic uncertainty on the cross section is
listed in Table~\ref{tabl}.

\section{Conclusion}
In the experiment at the VEPP-2000 $e^+e^-$ collider with the SND detector,
the cross section of the $e^+e^-\to\eta\gamma$ process was measured in the
energy range 1.05 -- 2.00 GeV. $\eta\gamma$ events were searched in the
$\eta$ decay mode $\eta\to 3\pi^0\to 6\gamma$.
%In this work, we have analyzed the data set accumulated in the experiment with
%the SND detector at the $e^+e^-$ collider VEPP-2000 in the center-of-mass
%energy range from 1.05 to 2.00 GeV. In the seven-photon final state, 
%events of the $e^+e^- \to \eta\gamma$ process have been selected.
The measured cross section of this process is shown in Fig.~\ref{fig:crs}
in comparison with the previous SND result~\cite{SNDetagam2014} obtained using
approximately 7 times less statistics. The new results are significantly
lower than the previous ones for $\sqrt{s}>1.25$ GeV. The difference is
explained by a significant underestimation of background in 
Ref.~\cite{SNDetagam2014}. The results obtained in this work supersede
the measurement of Ref.~\cite{SNDetagam2014}.

As a result of the fit to the cross section with the vector meson
dominance model, the values of the cross sections at the resonance maxima
have been obtained:
\begin{eqnarray}
\sigma_{\rho^{\prime}\to\eta\gamma} = 16^{+15}_{-10} \pm 2\mbox{ pb},\nonumber \\
\sigma_{\phi^{\prime}\to\eta\gamma} = 14^{+14}_{-10} \pm 2\mbox{ pb}, \nonumber
\end{eqnarray}
which agree with the estimates 
$\sigma_{\rho^{\prime}\eta\gamma} \approx 15$ pb, 
$\sigma_{\phi^{\prime}\eta\gamma} \approx 10$ pb 
made in Ref.~\cite{SNDetagam2014} basing on the quark-model prediction 
$\Gamma_{\rho^{\prime}\to\eta\gamma}\approx\Gamma_{\phi^{\prime}\to\eta\gamma}\approx 100$ keV~\cite{thpred}.

\end{document}